\documentclass[runningheads,a4paper]{llncs}
\usepackage{etex}
\reserveinserts{28}
\pdfoutput=1

\usepackage[linesnumbered,lined,boxed,commentsnumbered,ruled]{algorithm2e}

\SetAlFnt{\small}
\SetAlCapFnt{\small}
\SetAlCapNameFnt{\small}
\SetAlCapHSkip{0pt}
\IncMargin{-\parindent}
\usepackage{ textcomp }
\usepackage{ comment }
\usepackage[color=yellow!60,textsize=footnotesize,obeyDraft,draft]{todonotes}
\usepackage{colortbl}
\usepackage{booktabs}
\usepackage{capt-of}
\usepackage{makecell}
\usepackage{textcomp }
\usepackage{multirow}
\usepackage{listings}
\usepackage{graphicx}
\usepackage{colortbl}
\usepackage{booktabs}
\usepackage{amsmath}
\usepackage[font=footnotesize,labelfont=bf]{subcaption}
\usepackage[hidelinks]{hyperref}
\usepackage[capitalize]{cleveref}
\usepackage{tikz}
\usepackage{pgfplots}
\usepackage[para,online,flushleft]{threeparttable}
\usepackage{multirow}
\usepackage{wasysym}
\usepackage{amssymb}
\usepackage{enumitem}
\usetikzlibrary{arrows}
\usepackage{subcaption}
\usepackage{breakcites}
\newcounter{mnotei}
\newcolumntype{L}[1]{>{\raggedright\let\newline\\\arraybackslash\hspace{0pt}}m{#1}}
\newcolumntype{C}[1]{>{\centering\let\newline\\\arraybackslash\hspace{0pt}}m{#1}}
\newcolumntype{R}[1]{>{\raggedleft\let\newline\\\arraybackslash\hspace{0pt}}m{#1}}

\newcommand{\includegraphicsmaybe}[2]{
    \IfFileExists{#2}{\includegraphics[#1]{#2}}{
    \detokenize{File #2 is missing, maybe you need to run plots.py?}
}}

\begin{document}
\bibliographystyle{ieeetr}
\mainmatter

\title{Energy Transparency for Deeply Embedded Programs}

\titlerunning{Energy Transparency for Deeply Embedded Programs}

\author{Kyriakos Georgiou\inst{1}, Steve Kerrison\inst{1}, Zbigniew Chamski\inst{2}, Kerstin Eder\inst{1}}

\authorrunning{K. Georgiou et al.}
\institute{University of Bristol \and Infrasoft IT Solutions, Poland}
\tocauthor{Authors' Instructions}
\maketitle

\makeatletter
\renewcommand\subsubsection{\@startsection{subsubsection}{3}{\z@}%
                       {-18\p@ \@plus -4\p@ \@minus -4\p@}%
                       {4\p@ \@plus 2\p@ \@minus 2\p@}%
                       {\normalfont\normalsize\bfseries\boldmath
                        \rightskip=\z@ \@plus 8em\pretolerance=10000 }}
\makeatother

\begin{abstract}
Energy transparency is a concept that makes a program's energy consumption visible, from hardware up to software, through the different system layers.
Such transparency can enable energy optimizations at each layer and between layers, and help both programmers and operating systems make energy-aware decisions. 
In this paper, we focus on deeply embedded devices, typically used for Internet of Things (IoT) applications, and demonstrate how to enable energy transparency through existing Static Resource Analysis (SRA) techniques and a new target-agnostic profiling technique, without hardware energy measurements.
Our novel mapping technique enables software energy consumption estimations at a higher level than the Instruction Set Architecture (ISA), namely the LLVM Intermediate Representation (IR) level, and therefore introduces energy transparency directly to the LLVM optimizer.
We apply our energy estimation techniques to a comprehensive set of benchmarks, including single- and also multi-threaded embedded programs from two commonly used concurrency patterns, task farms and pipelines. 
Using SRA, our LLVM IR results demonstrate a high accuracy with a deviation in the range of 1\% from the ISA SRA. Our profiling technique captures the actual energy consumption at the LLVM IR level with an average error of 3\%.
\end{abstract}

\section{Introduction}
\label{sec:introduction}

The various abstraction layers introduced through the system stack to make programming easier, make it very difficult to understand how coding and data structures affect the energy consumption when the program is executed.
Energy transparency aims to leverage this information from the lower levels of the system stack up to the user~\cite{Eder:2016:ENTRA}. Such information can be of significant value for Internet of Things (IoT) applications which typically have to operate on limited or unreliable sources of energy. 

Deploying millions of embedded devices into IoT environments poses the challenge of how to power
them. Battery-based solutions can be costly and impractical due to the need for replacement. A
better solution is a combination of energy harvesting with ultra-low energy embedded devices. Energy harvesting comes with two caveats. Firstly, it is an unreliable source of energy and, secondly, it cannot yet deliver the required energy budgets for many IoT applications.
While many achievements have been made in optimizing the energy consumption of hardware, too little is done to expose potential hardware-level energy savings to the software developer. We
propose techniques for exposing the bounds and the actual energy consumption of software written
for a specific platform, as part of the embedded systems software development cycle. This enables programmers, tool chains and runtime systems to make energy-aware decisions in order to meet their strict energy constraints.

The energy consumption of a program on specific hardware can always be determined through physical measurements. Although this is potentially the most accurate method, it is often not easily accessible. Measuring energy consumption can involve sophisticated equipment and special hardware knowledge. Custom modifications may be needed to probe the power supply. Even though energy monitoring counters are becoming increasingly popular in modern processors, their number and availability are still limited in deeply embedded systems. Moreover, fine-grained energy characterization of software components, such as Control Flow Graph (CFG) basic blocks or loops, can not be achieved by just using energy measurements. All this makes it very difficult for the majority of software developers to assess a program's energy consumption.

Energy consumption is a resource constraint, with another frequently examined constraint being
execution time. Significant progress has been made in the area of Worst Case Execution Time (WCET) prediction using Static Resource Analysis (SRA) techniques that determine safe upper bounds for the execution time of programs. 
A popular approach used for WCET is the Implicit Path Enumeration Technique (IPET), which retrieves the worst case control flow path of programs based on a timing cost model.
Instead, in \cite{Jayaseelan2006}, an energy model that assigns energy values to blocks of Instruction Set Architecture (ISA) code is used to statically estimate Worst Case Energy Consumption (WCEC). We also adopted IPET in our SRA, retrieving energy bounds for the processor under investigation, the XMOS XS1-L ``Xcore''.

The Xcore is a multi-threaded deeply embedded processor with time deterministic instruction
execution. Such systems are simpler than general purpose processors and favor predictability and low energy consumption over maximizing performance. Processors with such characteristics are the
backbone for IoT applications. Moreover, the absence of performance-enhancing complexity at the
hardware level, such as caches, make them ideal for critical applications. We base our choice of the Xcore processor among other more popular architectures used for IoT applications, such as the
\texttt{ARM Cortex-M} series~\cite{Cortex_M_Series}, on the fact that the Xcore is a time deterministic multi-threaded architecture that can be extended to many-core systems~\cite{Kerrisson:2016}, allowing for a variety of design space exploration choices. Our SRA uses an ISA multi-threaded energy model for the Xcore introduced in \cite{Kerrison15}. 

In addition, we have developed a novel mapping technique to lift our ISA-level energy model to a higher level, the intermediate representation of the compiler, namely LLVM IR~\cite{LattnerLLVM2004}, implemented within the LLVM tool chain~\cite{LLVM}.
This enables SRA to be performed at a higher abstraction level than ISA, thus introducing energy transparency into the compiler tool chain by making energy consumption information accessible directly to the optimizer. Transparency of energy consumption at this level enables programmers to investigate how optimizations affect their program's energy consumption~\cite{TLP:9940890}, or even helps introduce new low energy optimizations~\cite{7054192,Pallister:2014}. This is more applicable at the LLVM IR than at the ISA level, because more program information exists at that level, such as types and loop structures. Our mapping and analysis techniques at the LLVM IR level are applicable to any compiler that uses the LLVM common optimizer, provided that an energy model for the target architecture is available. Our LLVM IR analysis results demonstrate a high accuracy with a deviation in the range of 1\% from the ISA SRA.

Because currently, there is no practical method to perform actual case static
analysis~\cite{townley2013practical}, we introduce a profiling technique that can provide actual
energy estimations directly at the LLVM IR level. The profiler is implemented at the LLVM IR level in order to keep the technique as target-agnostic as possible. Moreover, the technique is more favorable than Instruction Set Simulation (ISS) based estimations, as building an ISS for an architecture is a significantly bigger task. Our profiling-based estimation can provide at least the same performance or significantly outperform the estimation speed of an ISS-based estimation, depending on the complexity of the algorithm implemented and the size of the resulting program. This makes it more suitable for iterative optimizations during software development. The profiling-based estimation is evaluated on a large set of benchmarks, showing an only 1.8\% deviation compared to a cycle accurate ISS for the Xcore.

The main contributions of this paper are:
\begin{enumerate}
\item Formalization and implementation of a novel target-agnostic mapping technique that lifts an ISA-level energy model to a higher level, the intermediate representation of the LLVM compiler (\Cref{sec:mapping});
\item SRA energy estimation at the ISA level and at the LLVM IR level using our mapping technique (\Cref{sec:SAnalysis});
\item A new target-agnostic profiling-based energy estimation technique, that retrieves estimations at the LLVM IR level by the use of our mapping technique, and with an accuracy close to a cycle accurate ISS-based estimation (\Cref{subsec:Profiling});
\item SRA and profiling extension for a set of multi-threaded programs (\Cref{subsec:mult-threadedAnalysis}), focusing on task farms and pipelines, two commonly used concurrency patterns;
\item Comprehensive evaluation of our SRA and profiling energy estimation techniques and our mapping technique accuracy on a large set of benchmarks (\Cref{sec:evaluation}).
\end{enumerate}

The rest of the paper is organized as follows. \Cref{sec:Xcore} gives an overview of the Xcore architecture and its ISA energy model. \Cref{sec:mapping} introduces our mapping technique and its instantiation for the Xcore processor. \Cref{sec:estimationTechs} details our two estimation methods, SRA and profiling-based. Our experimental evaluation methodology, benchmarks and results are presented and discussed in \Cref{sec:evaluation}. \Cref{sec:background} critically reviews previous work related to ours. Finally, \Cref{sec:conc_future} concludes the paper and outlines opportunities for future work.

\section{Xcore Architecture and Energy Model}
\label{sec:Xcore}

\subsection{Xcore Architecture Overview}

The Xcore processor is a deeply embedded processor intended to allow hardware interfaces to be implemented in software. This makes the Xcore well suited to embedded applications requiring
multiple hardware interfaces with real-time responsiveness. Interfaces, such as SPI, I2C and USB, can be written efficiently in C-style software, rather than relying on hardware blocks provided in a System-on-Chip or synthesized onto FPGA. To achieve this, the Xcore is designed to be a time deterministic hardware multi-threaded architecture, that provides inter-thread communication and I/O port control directly in the ISA. Moreover, for energy efficiency the processor is event-driven; busy waiting is avoided in favor of hardware-scheduled idle periods.

The processor supports up to 8 threads, with machine instructions and hardware resources dedicated
to thread creation, synchronization and destruction. Each thread has its own instruction buffer and
 register bank. The pipeline of the processor was designed to provide time-deterministic
execution and to maximize responsiveness, making the processor ideal for IoT applications. It is integer only, with no floating point hardware. By design the architecture avoids the need for forwarding between pipeline stages, speculative instruction issue and branch prediction and allows for zero time overhead in thread context switching. Threads are executed round-robin through a four-stage pipeline. Each thread can have only one instruction occupy the pipeline at any time, avoiding data hazards. Therefore, if fewer than four threads are active there will be clock cycles with inactive pipeline stages. This means that, to reach the maximum computation power of the processor, four or more threads have to be active to fully occupy the processor's pipeline. When more than four threads are active, maximum throughput is maintained, but compute time is divided between active threads.

The Xcore is many-core scalable, with 2- or 5-wire ``X-Links'' and a network
switch embedded into each core. Communication between threads on the same or
different cores is done via synchronous channel-based message passing. Threads
on the same core can communicate without contention, whereas the links used for
multi-core communication may be contended. These properties allow design space
exploration of multi-threaded programs that are core-local and/or multi-core. A
case of the trade-off between processor count and energy efficient execution is
investigated in~\Cref{subsubsect:multi-threaded_usage}. Full details of the
architecture are provided in \cite{xs1architecture}.

\subsection{Xcore Multi-threaded ISA Energy Model}
\label{subsec:xs1model}

The underlying energy model for this work is captured at the ISA level. Individual instructions from the ISA are assigned a single cost each. These can then be used to compute power or energy for sequences of instructions. The modeling technique is built upon \cite{Tiwari1996}, which is adapted and extended to consider the scheduling behavior and pipeline characteristics of the Xcore~\cite{Kerrison15}. The model captures the cost of thread scheduling performed by the hardware, in accordance with a series of profiling tests and measurements, because it influences the energy consumption of program execution. The cost associated with an instruction represents the average energy consumption obtained from measuring the energy consumed during instruction execution based on constrained pseudo-randomly generated operands.

A new version of this model that is well suited for static analysis has been developed. It
represents energy in terms of static and dynamic power components to better reflect
inter-instruction and inter-thread overheads. This has improved model accuracy by an average of four percentage points.

\begin{equation}
  E_{prg} =  \left(P_s + P_{di}\right) \cdot T_{idl} + \sum_{i \in
        {prg}}\left(
          \frac{P_s + P_i M_{N_p} O}{N_p} \cdot 4 \cdot T_{clk}
        \right)
\text{, where } N_p = \min(N_t, 4)
   \label{eq:xs1model_new_mt}
\end{equation}

\indent In \Cref{eq:xs1model_new_mt}, $E_{prg}$ is the energy of a program, formed by adding the energy consumed at idle to the energy consumed by every instruction, $i$, executed in the program. At idle, only a base processor power, the sum of its static power, $P_s$, and dynamic idle power, $P_{di}$, is dissipated for the total idle time, $T_{idl}$. For each instruction, static power is again considered, with additional dynamic power for each particular instruction, $P_i$. The dynamic power contribution is then multiplied by a constant inter-instruction overhead, $O$, that has been established as the average overhead of instruction interleaving. This is then
multiplied by a scaling factor to account for the number of threads in the pipeline, $M_{N_p}$. The result is divided by the number of instructions in the pipeline, which is at most four and is dependent upon the number of active threads, $N_t$. Each instruction completes in four cycles, so $4 \cdot T_{clk}$ gives the energy contribution of the given instruction, based on the calculated power.

When more than four threads are active, the issue rate of instructions per thread will be reduced. The energy model accounts for this with the $\min$ term in \Cref{eq:xs1model_new_mt}. From a purely timing perspective, the latency between instruction issues for a thread is $\max\left(N_t,4\right) \cdot T_{clk}$. This property means that instructions are time-deterministic, provided the number of active threads is known. A thread may stall in order to fetch the next instruction. This is also deterministic and can be statically identified~\cite[pp.\  8--10]{xs1architecture}. These instruction timing rules have been used in simulation-based energy estimation, and are also utilized in both the SRA and profiling performed in this paper.

A limited number of instructions can be exceptions to these timing rules. The divide and remainder instructions are bit-serial and take up to 32 cycles to complete. Resource instructions may block if a condition of their execution is not met, e.g.\ waiting on inbound communication causes the instruction's thread to be de-scheduled until the condition becomes satisfied. This paper focuses its contributions on fully predictable instructions, with timing disturbances from communication forming future work.

\subsection{Utilizing the Xcore energy model for energy consumption estimations}
\label{subsub:utilizingEM}

To determine the energy consumption of a program based on \Cref{eq:xs1model_new_mt} the  program's instruction sequence, $\langle i_1, \dots, i_n\rangle$, the idle time $T_{idl}$, and the number of active threads $N_p$ during instruction execution must be known.
In \cite{Kerrison15} an ISS was used to gather full trace data or execution statistics to obtain these parameters. In this work we use ISS only as a reference for comparison of SRA and profiling results, with a second reference being direct hardware measurement. 

For both SRA and profiling, we need to extract the CFGs for each thread and identify the interleavings between them. This allows for each instruction in the program to identify the $N_p$ component in \Cref{eq:xs1model_new_mt}. It also allows to estimate the total idle time, $T_{idl}$ of the program. 
For single-threaded programs the energy characterization of the CFG is straightforward, as there is no thread interleaving. For SRA, the IPET can be directly applied to the energy characterized CFG to extract a path that bounds the energy consumption of the program, as described in \Cref{sec:SAnalysis}. For arbitrary multi-threaded programs, using static analysis to energy characterize the CFG of each thread is challenging. We have therefore concentrated on two commonly used concurrency patterns, task farms and pipelines, which we use with evenly distributed workloads across threads. For these classes of programs the number of active threads across the whole execution is constant and equal to the number of threads used to implement the task farm or the pipeline. Similarly, the profiling technique does not need to account for thread interleavings for the programs investigated in this work. 

In addition to instructions defined in the ISA, a Fetch No-Op (FNOP) can also be issued by the processor. These occur deterministically~\cite[pp.\ 8--10]{xs1architecture}. FNOPs can significantly impact on energy consumption, particularly within loops. To account for FNOPs in both our SRA and profiling, the program's CFG at ISA level is analyzed. An instruction buffer model is used to determine where FNOPs will occur in a basic block (BB). Further details on FNOPs modeling can be found in \cite{fnops}.

\section{Mapping ISA Code to LLVM IR and LLVM IR Energy Characterization}
\label{sec:mapping}

Our mapping technique aims to link each LLVM IR instruction of a program with its corresponding machine specific ISA emitted instructions. Such a mapping can give powerful insights to the LLVM IR optimizer regarding code size, execution time and the energy consumption of a program. Furthermore, our LLVM mapping technique does not involve statistical analysis; instead, it is an on-the-fly technique that takes into consideration the compiler behavior and the actual program's CFG structure. The technique is fully portable and target agnostic. It requires only the adjustment of the LLVM mapping pass to the new architecture.

In this section we first formalize our generic mapping technique. We then specialize the technique to determine the energy characteristics of LLVM IR instructions. This specialization propagates ISA level energy models up to LLVM IR level, enabling energy consumption estimation of programs at that level. Finally, we instantiate and tune the mapping technique for the architecture under consideration, the Xcore.

\subsection{Formal specification of the mapping}
\label{mappingFormal}

The main idea of the mapping technique is to monitor the back-end of a compiler to establish a $1:m$ relation between the optimized LLVM IR and the emitted ISA. The goal of this mapping is to associate a single LLVM IR instruction with all the ISA instructions that originated from it, whenever possible. Then, by aggregating the energy costs of these ISA instructions, we can assign an energy cost to their single corresponding LLVM IR instruction. The mapping technique also guarantees that there is no loss of energy between the two levels as each ISA instruction will be mapped to one LLVM IR instruction. We formalize the mapping as follows. For a program $P$, let

\begin{equation}
    \mathrm{IRprog}_L = \{1,2,...,n\}
\label{eq:LLVMIDs}
\end{equation}
be the ID numbers of $P$'s LLVM IR instructions after the LLVM transformation and optimizations passes, with
\begin{equation}
  \mathrm{IRprog}=\langle ir_{1},ir_{2},...,ir_{n}\rangle
\label{eq:LLVMIRcode}
\end{equation}
being the sequence of LLVM IR instructions for $P$.
An architecture-specific compiler back-end $T_{arch}$ translates the IR program into ISA code:
\begin{equation}
 \mathrm{T}_{arch}(\mathrm{IRprog}) = \mathrm{ISAprog} = \langle isa_{1},isa_{2},...,isa_{k}\rangle\label{eq:tarch}
\end{equation}
producing a sequence of machine instructions $\langle isa_{1},isa_{2},...,isa_{k}\rangle$ that represents the program $P$ at ISA level. Let
\begin{equation}
\begin{split}
& \mathrm{M}(\mathrm{IRprog})=\langle (isa_{1},m_1),(isa_{2},m_2),...,(isa_{k},m_l) \rangle \\ 
& \text{ where } m_1, m_2,...,m_l \in \mathrm{IRprog}_L
\end{split}
  \label{eq:ISAcode}
\end{equation}
be the mapping process that monitors $\mathrm{T}_{arch}$ and creates a relation between the sequence of ISA instructions for $P$ and the IDs of the LLVM IR instructions, with the aim to associate an ISA instruction with the LLVM IR instruction it originated from, whenever this is possible. For ISA instructions that are not related to any LLVM IR instruction (ISA injected instructions) or whose origin LLVM IR instruction is ambiguous, the mapping process has to make an implementation-specific choice as to which LLVM IR instruction an ISA instruction should be associated with. This will preserve the 1:m relation and ensure that such instructions are accounted at the LLVM IR level. The mapping function

\begin{equation}
\begin{split}
& \mathrm{R}(ir_i) = \{ isa_j | ir_i \in \mathrm{IRprog} \wedge isa_j \in \mathrm{ISAprog} \wedge (isa_{j},i) \in \mathrm{M}(\mathrm{IRprog}) \}  \text{ with } \\
& \text{~~~~~the property } \forall ~ ir_n, ir_k \in \mathrm{IRprog} \wedge n \neq k \text{ then }
 \mathrm{R}(ir_n) \cap \mathrm{R}(ir_k) = \emptyset
\end{split}
\label{eq:mapping}
\end{equation}
captures a $1:m$ relation from $\mathrm{IRprog}$ to $\mathrm{ISAprog}$ instructions. The energy consumption of an LLVM IR instruction can be retrieved by
 \begin{equation}
   \mathrm{E}(ir_{i}) = \sum_{isa_j \in S} \mathrm{E}(isa_{j}) \text{ where } ir_{i} \in \mathrm{IRprog} \wedge isa_{j} \in \mathrm{ISAprog} \wedge \mathrm{S}=\mathrm{R}(ir_i)
 \label{eq:energy}
\end{equation}
as the sum of the energy consumed by all ISA instructions mapped to that LLVM IR instruction. \Cref{eq:energy} can also be used to associate timing or code size information with LLVM IR instructions, by replacing the ISA instructions' energy costings with the resource of interest costings.

\subsection{Xcore mapping instantiation and tuning}
\label{subsec:mapXcore}

\begin{figure}
        \centering
        \includegraphics[width=1.14\textwidth,clip,trim=0.12cm 6.3cm 1cm 7.2cm]{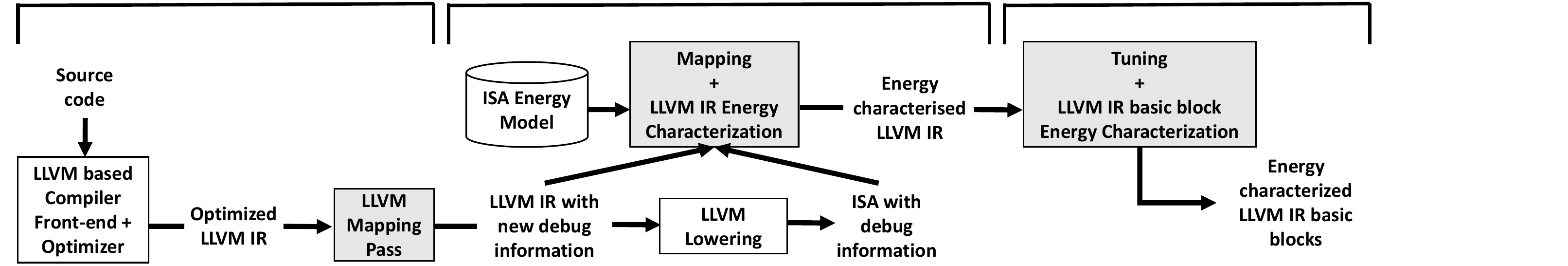}%
        \caption{LLVM IR energy characterization overview.}
        \label{fig:mappingOverview}
\end{figure}

In our case, the $\mathrm{T}_{arch}$ function is the XMOS tool chain lowering phase that translates
the LLVM IR to Xcore-specific ISA. The real challenge is to create from scratch a mechanism that can monitor $\mathrm{T}_{arch}$ and create the mapping given by \Cref{eq:ISAcode} which will have the property of \Cref{eq:mapping}. This would require an extensive knowledge of the back-end of the
architecture under consideration and a vast engineering effort to take into account every possible
transformation and optimization that happens between the optimized LLVM IR and the final ISA emitted code. Instead, we demonstrate a simple yet powerful technique that can provide sufficiently accurate results. An overview of the technique is given in \Cref{fig:mappingOverview}. We now describe the three mapping stages.

\subsubsection{Stage 1: LLVM IR annotation}
\label{subsub:LLVMAnnotation}

The goal of this stage is to enable the property of~\Cref{eq:mapping} that ensures a $1\!:\!m$ relation between the LLVM IR and ISA code. To achieve this, our mapping implementation leverages the debug mechanism in the XMOS compiler tool chain. Symbols are created during compilation to assist with debugging. These symbols are propagated to all intermediate code layers and down to the ISA code. Debug symbols can express which programming language constructs generated a specific piece of machine code in a given executable module. In our case, these symbols are generated by the front-end of the XMOS compiler in standard DWARF format~\cite{DWARF:2013:Online}. These are transformed to LLVM metadata~\cite{LLVMmetaData:2014:Online} and attached to the LLVM IR. LLVM 2.7 and upwards uses this metadata format as the primary means of storing debug information.

During the lowering phase of compilation, LLVM IR code is transformed to a target ISA by the back-end of the compiler, with debug information stored alongside in the DWARF standard format.
Tracking source code debug information gives an $n\!:\!m$ relationship between instructions at the different layers, because several source code instructions can be translated to many LLVM IR instructions, and these again into many ISA instructions. This $n\!:\!m$ relation prevents the fine-grained energy mapping needed for accurate energy estimations.

To achieve a $1\!:\!m$ mapping between the optimized LLVM IR instructions and the ISA instructions using the debug mechanism, we created an LLVM pass that traverses the optimized LLVM IR and replaces source location information with LLVM IR location information, or adds new location information to LLVM IR instructions without any debug data. This LLVM pass runs after all optimization passes, just before emitting ISA code, because the optimized LLVM IR is closer in structure to the ISA code than the unoptimized version. An example output of this process is given on the left hand side of \Cref{fig:mapping}. This represents a part of the LLVM IR CFG of a program after the LLVM optimizations, along with the unique debug location, $\mathrm{IRprog}_L$ in \Cref{eq:LLVMIDs}, assigned to each LLVM IR instruction.

The XMOS compiler debug mechanism applies the following four rules to preserve the debug information through the different transformations and optimizations applied in the back-end:

\begin{enumerate}
\item If an instruction is eliminated, its debug location information is also eliminated. 
\item If one instruction is transformed into another one, the debug location of the original instruction is assigned to the new instruction.
\item If multiple instructions are merged into one, the debug location of one of the initial instructions, the first one in our case, is assigned to the new instruction.
\item If one instruction is transformed into multiple ones, then all the new instructions are being assigned the debug location of their origin instruction.
\end{enumerate}

These rules iteratively preserve the $1\!:\!m$ relationship between the LLVM IR instructions in $\mathrm{IRprog}$ and the final ISA in $\mathrm{ISAprog}$, with two exceptions. 

The first is when instructions are introduced at the ISA level and they do not correspond to any LLVM IR instruction (ISA injected instructions). In such cases the mapping process can assign to them the debug location of an adjacent ISA instruction in the same BB. This ensures that they are accounted for in the mapped LLVM IR block. The second is when a transformation takes as an input multiple instructions and converts them to another set of multiple instructions in a single step. An example of such transformations are peephole optimizations. In such cases the handling of debug information depends on the compiler implementation. For the back-end under investigation, ISA instructions generated by such transformations are left without any debug information.
To account for these instructions at the LLVM IR level, we use the same approach as for ISA injected instructions.

\subsubsection{Stage 2: Mapping and LLVM IR Energy Characterization}
\label{subsub:MapChar}

Once the LLVM IR annotation has been performed for a program, the back-end lowering phase translates the LLVM IR code to target-specific ISA code. Then the mapping phase, which implements~\Cref{eq:mapping}, runs and maps LLVM IR instructions with the new debug locations to the emitted ISA instructions which carry the same debug location ID. Finally, the energy values for groups of ISA instructions are aggregated and then associated with their single corresponding LLVM IR instruction, as described by \Cref{eq:energy}.

\begin{figure}
        \centering
        \includegraphics[width=1\textwidth,clip,trim=0.38cm 7.5cm 1cm 3.5cm]{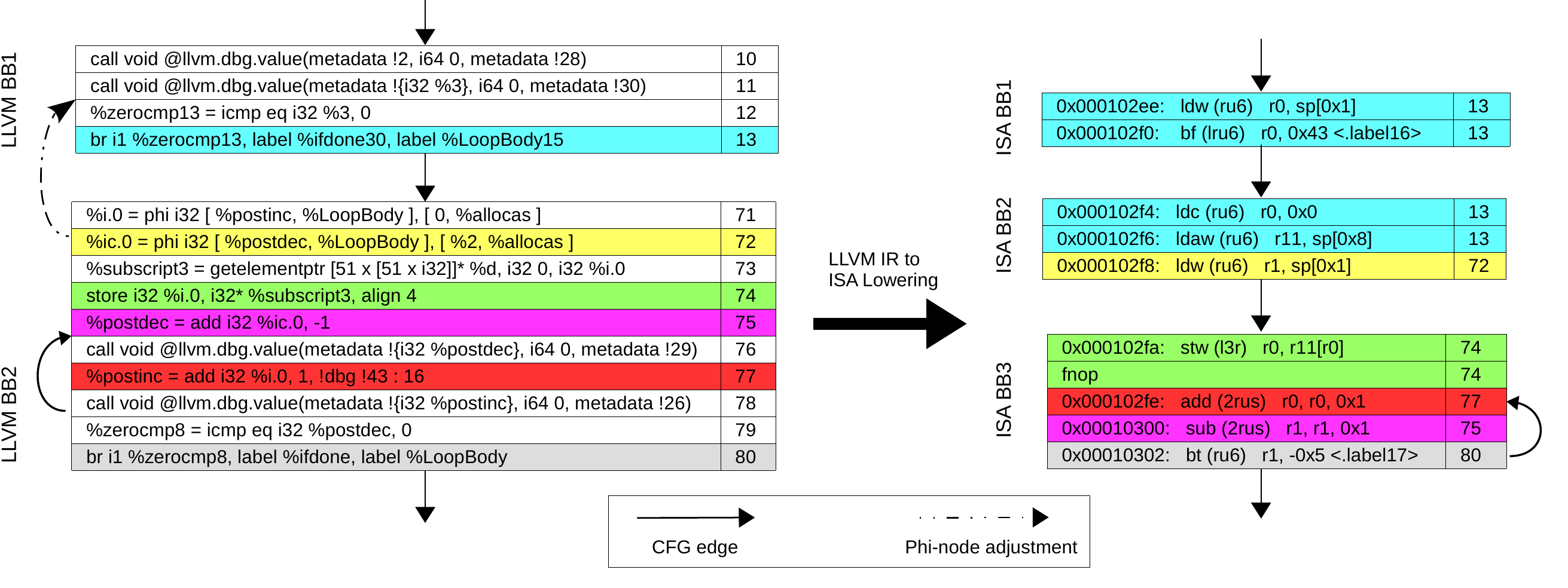}%
        \caption{Fine-grained $1\!:\!m$ mapping including our LLVM mapping pass.}
        \label{fig:mapping}
\end{figure}

An example mapping is given in \Cref{fig:mapping}. On the left-hand side is a part of the LLVM IR CFG of a program, which represents the $\mathrm{IRprog}$ in \Cref{eq:LLVMIRcode}, along with the new debug location assigned to each LLVM IR instruction by the mapping pass and represented by \Cref{eq:LLVMIDs}. The right-hand side shows the corresponding ISA CFG, together with the debug locations for each ISA instruction, given by \Cref{eq:ISAcode}. The coloring of the instructions demonstrates the mapping between the two CFGs' instructions using \Cref{eq:mapping}. Now, one LLVM IR instruction is associated with many ISA instructions, but each ISA instruction is mapped to only one LLVM IR instruction. Some LLVM IR instructions are not mapped, because they are removed during the lowering phase of the compiler. This mapping also guarantees that all ISA instructions are mapped to the LLVM IR, so there is no loss of energy between the two levels.

This Xcore instantiation of the mapping technique, using the debug mechanism, can be ported to any architecture supported by the LLVM IR compiler for which an ISA-level energy model exists. Typically, these are deeply embedded processors, such as the one used here and the \texttt{ARM Cortex M} series, which are ideal for IoT applications. 

\subsubsection{Stage 3: Tuning and Basic-Block Energy Characterization}
\label{subsub:MapTuning}

Without any further tuning, the mapping instantiation for the Xcore architecture provides an average deviation of 6\% between the SRA prediction at the LLVM IR level and that at the ISA level. An additional tuning phase is introduced after the mapping, to account for specific architecture behavior and to facilitate our BB level energy analysis. In the case of the bound analysis, this improved the LLVM IR SRA accuracy, narrowing the gap between ISA and LLVM IR energy predictions to an average of 1\% as demonstrated in our results in~\Cref{sec:BSAresults}. This tuning had a similar positive effect on the accuracy of the LLVM IR profiling-based energy estimation. 

As discussed in \Cref{subsub:utilizingEM}, FNOPs can be issued by the processor and this can be statically determined at the ISA level. This can not be represented in LLVM IR. Ignoring FNOPs can therefore lead to a significant underestimation of energy at the LLVM IR level. To account for FNOPs at the LLVM IR level, we treat them similarly to the ISA injected instructions; by assigning them the debug location of an adjacent ISA instruction in the same BB. \Cref{fig:mapping}, provides an example of such FNOP treatment at debug location number 74.

To estimate energy at the LLVM BB level, the energy cost of each BB is needed. 
This can be obtained by accumulating the energy costs of all LLVM IR instructions in an LLVM IR BB. When estimating energy at the BB level, the position of the LLVM IR instructions in BBs is critical, and tuning may be needed to transfer mapped energy costs to different BBs to reflect more accurately where the energy is consumed during program execution. \emph{Phi-nodes}, for example, benefit from such tuning.

\emph{Phi-nodes} can be introduced at the start of a BB as a side effect of the Single Static Assignment (SSA) used for variables in the LLVM IR. A \emph{phi-node} takes a list of pairs, where each pair contains a reference to the predecessor block together with the variable that is propagated from there to the current block. The number of pairs is equal to the number of predecessor blocks of the current block. A \emph{phi-node} can create inaccuracies in the mapping when LLVM IR is lowered to ISA code that no longer supports SSA, because it can be hoisted out from its current block to the corresponding predecessor block at the ISA level. For blocks in loops this can lead to a significant estimation error. Such cases can be detected by examining the mapping. Similar inaccuracies can be introduced by branching LLVM IR instructions with multiple targets, if the ISA of the target processor supports only single-target branches.

\IncMargin{1em}
\begin{algorithm}
\newcommand\mycommfont[1]{\scriptsize\ttfamily\textcolor{blue}{#1}}
\SetCommentSty{mycommfont}

\footnotesize
\DontPrintSemicolon

\SetKwData{ISAmap}{ISAmap}
\SetKwData{IRBBloopDepth}{irBBloopDepth}
\SetKwData{ISABBloopDepth}{isaBBloopDepth}
\SetKwData{ToIRBB}{ToIRBB}
\SetKwData{FromIRBB}{FromIRBB}
\SetKwData{ISABB}{ISABB}
\SetKwData{ISAcost}{ISAcost}

\SetKwFunction{GetISAmap}{GetISAmap}
\SetKwFunction{GetInstrBBLoopDepth}{GetInstrBBLoopDepth}
\SetKwFunction{FindIRBB}{FindIRBB}
\SetKwFunction{GetISABB}{GetISABB}
\SetKwFunction{Cost}{GetCost}
\SetKwFunction{AddCost}{AddCostToBB}
\SetKwFunction{RemoveCost}{RemoveCostFromBB}
\SetKwFunction{GetCurrentIRBB}{GetCurrentIRBB}
\SetKwInOut{Input}{input}\SetKwInOut{Output}{output}

\Input{$\mathrm{IRprog}$ for a program $P$ as described by \Cref{eq:LLVMIRcode}}
\Input{$\mathrm{IRprogCFG}$, the CFG for $\mathrm{IRprog}$, with the total energy cost for each BB}
\Input{$\mathrm{ISAprogCFG}$, the CFG for the $\mathrm{ISAprog}$ given by \Cref{eq:tarch} with BB total energy info}
\Input{$M$, the $1:m$ mapping retrieved for $P$ using the mapping as described in this section}

\For{each $ir_n \in \mathrm{IRprog}$}{
  \If{$ir_n$ is a \emph{phi-node}}{
    \ISAmap $\leftarrow$ \GetISAmap{$ir_n$, $M$} \tcp*[h]{all the ISA instructions mapped to $ir_n$}\;
    \IRBBloopDepth $\leftarrow$ \GetInstrBBLoopDepth{$ir_n$} \tcp*[h]{degree of nested loops for $ir_n$'s BB}\;
    \FromIRBB $\leftarrow$ \GetCurrentIRBB($\mathrm{IRprogCFG}$, $ir_n$) \tcp*[h]{the $ir_n$'s BB}\;
    \For{each $isa_k \in$ \ISAmap}{
      \ISABBloopDepth $\leftarrow$ \GetInstrBBLoopDepth{$isa_k$} \tcp*[h]{degree of nested loops for $isa_k$'s BB}\;
      \If{\IRBBloopDepth $>$ \ISABBloopDepth}{ 
        \ISABB $\leftarrow$ \GetISABB($\mathrm{ISAprogCFG}$, $isa_k$) \tcp*[h]{the $isa_k$'s BB}\;
        \ToIRBB $\leftarrow$ \FindIRBB($\mathrm{IRprogCFG}$, \ISABB)\label{findBBFunc} \tcp*[h]{finds the IR predecessor BB of \FromIRBB that matches \ISABB by examining the mapping of their instructions}\;
        \If{\ToIRBB not NULL}{
        \ISAcost $\leftarrow$ \Cost($isa_k$) \tcp*[h]{the energy cost of $isa_k$ from our ISA energy model}\;
        \RemoveCost(\FromIRBB, \ISAcost)\tcp*[h]{remove cost from initial block}\;
        \AddCost(\ToIRBB, \ISAcost) \tcp*[h]{add cost to the chosen predecessor block}\;
        }
      }
    }
  }
}
\caption{\emph{phi-node} tuning}\label{phinodesTunning}
\end{algorithm}\DecMargin{1em}

\Cref{phinodesTunning} detects cases where the ISA energy values mapped to \emph{phi-nodes} in a looping BB are accumulated into the wrong LLVM IR BB. It then hoist these costs out to the appropriate LLVM IR predecessor BBs. The algorithm detects the problematic cases by using the mapping and the BB loop depth. Then, by examining the mapping, function \texttt{FindIRBB} at \Cref{findBBFunc} finds the LLVM IR predecessor BB that matches the ISA block holding the ISA \emph{phi-node} generated instruction. The ISA's instruction energy cost is then added to the cost of the selected predecessor LLVM IR BB and subtracted from the \emph{phi-node's} LLVM IR BB.
Performing the mapping after the LLVM IR optimization passes and just before the lowering phase increases the possibility of similar CFG structures between the two levels. This improves the ability of \texttt{FindIRBB} to find the correct BB and therefore improves the results of the \emph{phi-node} tuning.

An example of a \emph{phi-node} adjustment is given in \Cref{fig:mapping} at debug location 72. Its corresponding ISA instruction is hoisted out from the loop BB, \texttt{ISA BB3}, and into
\texttt{ISA BB2}. A similar hoisting is performed from \texttt{LLVM BB2} and into \texttt{LLVM BB1} by \Cref{phinodesTunning}, thus correctly assigning energy values to each LLVM IR block.

\subsection{Limitations}
\label{subsec:limitations}

Our mapping approach between the LLVM IR and the ISA guarantees that no energy is lost between the two levels, as all the ISA instruction energy costs are propagated to the LLVM IR level. Therefore, any inaccuracies introduced in our LLVM IR energy estimations from the mapping are a consequence of attributing ISA energy costs to the wrong LLVM IR BBs. This is because our LLVM IR estimation techniques work at a BB level. In that respect, two cases can affect the estimation accuracy.

In the first case, for instructions at the boundaries of LLVM IR BBs, their corresponding ISA instructions may cross these boundaries at the ISA level. The mapping will, however, still associate the energy costs of these ISA instructions with their original LLVM IR instruction, and thus with their original LLVM IR BB. If such LLVM IR instructions belong to a BB that is part of a loop, when the ISA instruction has been hoisted out of that loop, then, with no proper adjustment, an overestimation will occur due to the mapping. An example of such a case are the Phi-node instructions, described in \Cref{subsub:MapTuning}, where Algorithm 1 is introduced to adjust their mapping. Such cases can be statically identified by examining the mapping. In the second case, a difference between the two CFG structures can occur when BBs are introduced/eliminated at the ISA level. If this makes the structures of the two CFGs significantly different, then the energy costs allocated to the LLVM IR BBs by the mapping can be inaccurate. Performing the mapping after the LLVM IR optimization passes and just before the lowering phase increases the possibility of having similar CFG structures between the two levels and therefore the accuracy of the mapping. The impact of the above issues is investigated in \Cref{sec:evaluation}.

\subsection{Discussion}

The mapping technique uses an ISA resource model. This has significant benefits over a stand alone LLVM IR static energy model. Firstly, our mapping-based approach benefits from the accuracy that ISA models can provide, because the ISA is closer to the hardware than LLVM IR. Secondly, the dynamic nature of the mapping technique can account for specific architecture behavior, such as the LLVM IR location to which the costs of the FNOPs should be attributed, and compiler specific behavior, such as code transformations. Static IR energy models that are created through statistical approaches are inherently limited in their ability to account for these. 

Furthermore, since \cite{Tiwari1996}, the seminal approach of constructing ISA energy models, there are several well-defined ISA energy models~\cite{Sarta1999,Brooks2000,Steinke2001,Sami2002,Ibrahim2008}. These models could now be lifted to the compilers' IR by our mapping. Our approach therefore benefits from a well understood process by which energy models can be created for deeply embedded systems, where predictability is provided at the ISA level.

\section{Energy Estimation Methods used}
\label{sec:estimationTechs}

\subsection{Static Resource Analysis}
\label{sec:SAnalysis}

Our IPET-based SRA is implemented in three stages detailed as follows: 

\begin{enumerate}
\item \textbf{Low-Level Analysis}: This stage aims to model the dynamic behavior of the processor's micro-architecture. For our energy consumption analysis, this is achieved through the ISA-level energy model detailed in \Cref{subsec:xs1model}, that captures the behavior of the Xcore processor with regard to its energy consumption characteristics. For the LLVM IR analysis an extra step is required to characterize the energy consumed by LLVM IR instructions as detailed in \Cref{subsec:mapXcore}.

\item \textbf{Control Flow Analysis}: This stage aims to capture the dynamic behavior of the program and associates CFG BBs with the information needed for the computation step of the analysis. IPET requires the CFG and call graph of a program to be constructed at the same level as the analysis. At LLVM IR level, the compiler can generate them. At ISA level a tool was created to construct them. To detect BBs that belong to a loop or recursion, we adopted and extended the algorithm in \cite{Wei:2007}. The CFGs are annotated according to the needs of the IPET
described in \cite{Li1997}. Finally, the annotated CFGs are used in the computation step to produce Integer Linear Programming (ILP) formulations and constraints.

\item \textbf{Computation}: The IPET adopted in our work to estimate the energy consumption of a program is based on \cite{Li1997}. To infer the energy consumption of a program, instead of using the time cost of a CFG BB, the BB's energy cost is used. The minimum required user input to enable bounding of the problem is the loop bounds declaration. This is also standard practice in timing analysis~\cite{Wilhelm:2008}. Further constraints, such as denoting CFG infeasible paths, can be provided in the form of user source code annotations, to extract more accurate estimations. The annotation language used in this work can be found in \cite{entra-d2.1}.
\end{enumerate}

\subsection{Profiling-Based Energy Consumption Estimation}
\label{subsec:Profiling}

In contrast to SRA, given a specific set of input parameters, our profiling technique aims to provide the actual energy consumed by a program, rather than bounds. To achieve this, the technique collects LLVM IR BB execution counts. The technique is target-agnostic in the sense that the instrumentation code is inserted at the architecture-independent LLVM IR level and does not require the modification of a program's object code to insert instrumentation instructions, unlike in other approaches~\cite{Srivastava:1994}. The energy consumption estimations are also collected at the LLVM IR level. This is enabled by the energy characterization of the LLVM IR code using the mapping technique introduced in \Cref{sec:mapping}.

\begin{figure}
        \centering
        \includegraphics[width=1.28\textwidth,clip,trim=0.12cm 3.9cm 1cm 8.8cm]{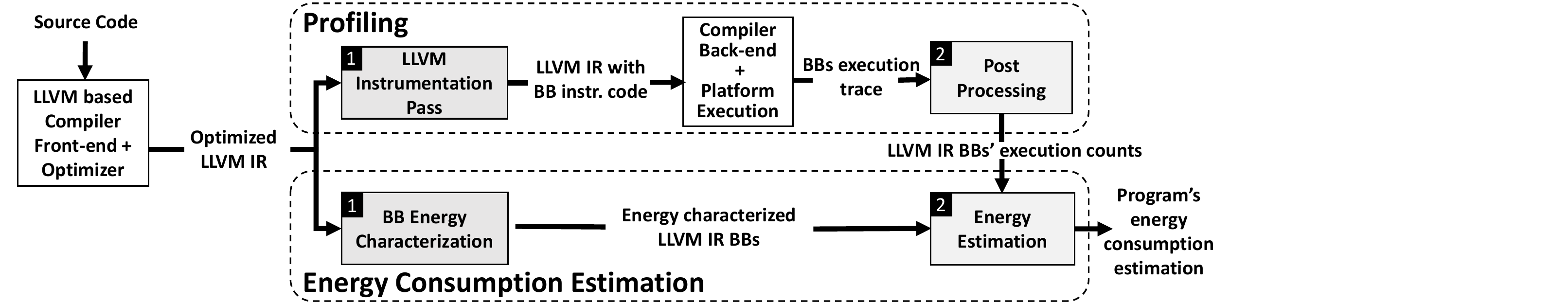}%
        \caption{Profiling-based energy estimation.}
        \label{fig:ProfilerOverview}
\end{figure}

\Cref{fig:ProfilerOverview} outlines the profiling-based energy consumption estimation process. The process is split in two phases: a profiling phase, which aims to collect BB execution traces, and an energy estimation phase that performs the LLVM IR energy characterization and combines it with the profiling information to obtain energy estimations. Both phases take as input the same LLVM IR, the one retrieved after all the optimization passes and just before emitting the target's ISA code. Using the optimized LLVM IR allows us to benefit from a more accurate energy characterization via the mapping technique, since it is closer in structure to the ISA code than the unoptimized version. This also avoids possible negative impact on the LLVM optimizations' abilities, resulting from the injected instrumentation code. The stages for each of the two phases are annotated with numbers in black boxes in \Cref{fig:ProfilerOverview} and are explained next.

\begin{itemize}[leftmargin=0cm,itemindent=0cm,labelwidth=\itemindent,labelsep=0cm,align=left]
\item[] \textbf{Profiling Phase}:
  \begin{enumerate}
  \item \textbf{~~LLVM Instrumentation Pass}: A new LLVM pass is built into the LLVM compiler that runs after all the LLVM optimization passes and just before the ISA lowering. The pass injects each LLVM IR BB with instrumentation instructions. To implement the BB instrumentation we considered two choices. First, using BB execution counters and second using instructions that emit a unique ID, identifying the BB and its origin function, every time the BB is executed. The second option puts less stress on the limited memory size, typically available on deeply embedded devices, as there is no need to keep global variables. Therefore, we chose to use the \emph{print} instruction supported in the LLVM IR assembly language. These \emph{print} instructions need to be translated by the compiler back-end to target-specific tracing functions. In the Xcore, \emph{printf} real-time debugging is supported by the \texttt{xTAG} debug adapter~\cite{xtag}, which emits the data to the host machine via buffering, with negligible impact on program execution time. The only requirement for other embedded devices to benefit from our profiler is the implementation of a target-specific instrumentation function to emit profiling data to the host machine, if not already in place. On most targets, this can be implemented using I/O ports, with very low impact on the program's execution time.
  \item \textbf{~~Post Processing}: The program is then executed and the BBs execution trace is collected. Based on the unique IDs emitted, the post-processing stage can associate execution counts with each BB of the optimized LLVM IR code.
  \end{enumerate}
\item[] \textbf{Energy Consumption Estimation Phase}:
  \begin{enumerate}
  \item \textbf{~BB Energy Characterization}: A clean copy of the optimized LLVM IR code, with no instrumentation instructions, is energy-characterized using the BB energy characterization mapping techniques introduced in \Cref{sec:mapping}. This ensures that no energy overheads appear in our energy consumption estimations due to instrumentation code.
  \item \textbf{~~Energy Estimation}: This stage takes as input the BB energy-characterized LLVM IR and the BBs execution counts collected from the Profiling phase, and produces the program's total energy consumption estimation. Having the LLVM IR BBs' execution counts and the LLVM IR instruction energy costings, a fine-grained software energy characterization can be achieved, where we can attribute energy consumption to specific programming constructs, e.g. BBs, loops and functions.
  \end{enumerate}
\end{itemize}

\subsection{Analysis of multi-threaded programs}
\label{subsec:mult-threadedAnalysis}

In this paper we present the first steps towards energy consumption analysis of multi-threaded programs. Two concurrency patterns are considered: replicated threads with no inter-thread communication, working on different sets of data (task farms), and pipelines of communicating threads. For both cases we consider evenly distributed, balanced workloads.

There is a fundamental difference when statically predicting the case of interest (worst, best, average case) for time and for energy for multi-threaded programs. Generally, for time only the computations that contribute to the path forming the case of interest must be considered. For energy, all computations taking place during the case of interest must be considered. For instance, in an unbalanced task farm, the WCET will be equivalent to the longest running thread. To bound energy, the energy consumed by each thread needs to be aggregated. Thus, the static analysis needs to determine the number of active threads at each point in time in order to apply the energy model from \Cref{eq:xs1model_new_mt} and characterize the CFG of each thread. Then, IPET can be applied to each thread's CFG, extracting energy consumption bounds. Aggregating these together will give a loose upper bound on the program's energy consumption, meaning that the safety of the bound cannot be guaranteed.

In our balanced task farm examples, all task threads are active in parallel for the duration of the test. Thus, the number of active threads is constant, giving a constant $N_t$, used to determine the pipeline occupancy scaling factor, $M$, in \Cref{eq:xs1model_new_mt}. For balanced pipelined programs we consider the continuous, streaming data use case, so the same constant thread count property holds. In both cases, IPET can be performed on each thread's CFG and the results aggregated to retrieve the total energy consumption. The energy model covers the core-local instructions used for thread communication. Although off-core communication uses the same instructions as core-local communication, an additional energy cost needs to be accounted for external-link usage. This cost was determined empirically for the development board used in \Cref{subsubsect:multi-threaded_usage}.

For multi-threaded programs with synchronous communications, to retrieve a WCET, IPET can be applied on a global graph, connecting the CFGs of all threads along communication edges. The communication edges can be treated by the IPET as normal CFG edges and WCET can be extracted by solving the formulated problem~\cite{PotopButucaru:2013}. This will return a single worst case path
across the global graph. Bounding energy in this way is not possible, as parallel thread activity over time needs to be considered. Activity can be blocked if the threads' workloads are unbalanced, due to the synchronous blocking message passing. In this case, statically determining the number of active threads at each point in time is hard.

Similar to SRA, profiling-based energy estimation does not need to infer the number of active threads at each execution stage, for our balanced task farms and pipelined test cases. Therefore, retrieving the BB execution counters is sufficient.  

The concurrency patterns addressed here represent typical embedded use cases. We demonstrate in \Cref{subsubsect:multi-threaded_usage} that for these use cases, SRA can provide sufficiently accurate information for design space exploration. More advanced concurrency analysis techniques, such as the ones employed in \cite{Shih:2014:COR}, could be combined with our techniques to scale our analysis to more complex concurrency patterns.

\section{Experimental Evaluation}
\label{sec:evaluation}

Two open source benchmark suites were used for evaluation.
The first one, \texttt{M{\"a}lardalen WCET} benchmark suite~\cite{Gustafsson2010}, is specially designed for WCET analysis. The second, the \texttt{BEEBS} benchmark suite~\cite{BEEBS}, is targeted at evaluating the energy consumption of embedded processors. The selected benchmarks were modified to work with our test harness and, in some cases, to make them more parametric to function input arguments. When necessary, benchmarks using floating point were modified to use integer values, since the Xcore does not support floating point operations. Some of the benchmarks were also extended to be multi-threaded task farms, where the same code runs on two or four threads. Furthermore, a range of industrial benchmarks was selected to demonstrate the value of our estimation techniques to embedded developers.   

Deeply embedded processors do not typically have hardware support for division or floating point operations, using software libraries instead. Software implementations are usually far less efficient than their hardware equivalent, both in terms of execution time and energy consumption. The effect of these software implementations on energy consumption should be known by developers, therefore we include software division and software-emulated floating point arithmetic benchmarks.
A radix-4 software divider, \texttt{Radix4Div}~\cite{Radix4Div}, is used. A less efficient version, \texttt{B.Radix4Div}, is added for comparison; it omits an early return when the dividend is greater than 255. As a consequence, CFG paths become more balanced, with less variation between the possible execution paths. The effect of this on the energy consumption is discussed later in this section. For software floating point, single precision \texttt{SFloatAdd32bit} and \texttt{SFloatSub32bit} operations from \cite{SoftFloat} are analyzed.

To extend our analysis to multi-threaded communicating programs, we analyze two common signal processing tasks written for the Xcore, \texttt{FIR} and \texttt{Biquad}~\cite{XMOSDSP}. 
Both tasks are implemented as pipelines of seven threads. Such programs are the preferred form for Xcore, as spreading the computation across threads allows the voltage and frequency of the core to be lowered, significantly reducing energy consumption while retaining the same performance as the single threaded version.

A complete list of all the benchmarks' attributes, can be found in \cite{benchmarks}. Benchmarks were compiled with \texttt{xcc} version 12~\cite{xtools} at optimization level \texttt{O2}; the default for most compilers. For hardware measurements, as in \cite{Kerrison15}, a shunt resistor current sense and sampling circuit obtains power dissipation with sub-milliwatt precision and variation of less than 1\,\%.

\subsection{Static Resource Analysis Results}
\label{sec:BSAresults}

For the evaluation of our SRA estimations together with the mapping technique, 20 benchmarks were used. These cover a broad spectrum of language features and code complexity. A combination of good understanding of the underlying algorithms, profiling information and brute forcing of the benchmarks' functions input space was necessary to identify tests covering the algorithmic worst case execution path for each benchmark with certainty.

\begin{figure*}[ht]
\centering
\includegraphics[width=1\textwidth,trim=0cm 0.45cm 0cm 0.4cm,clip]{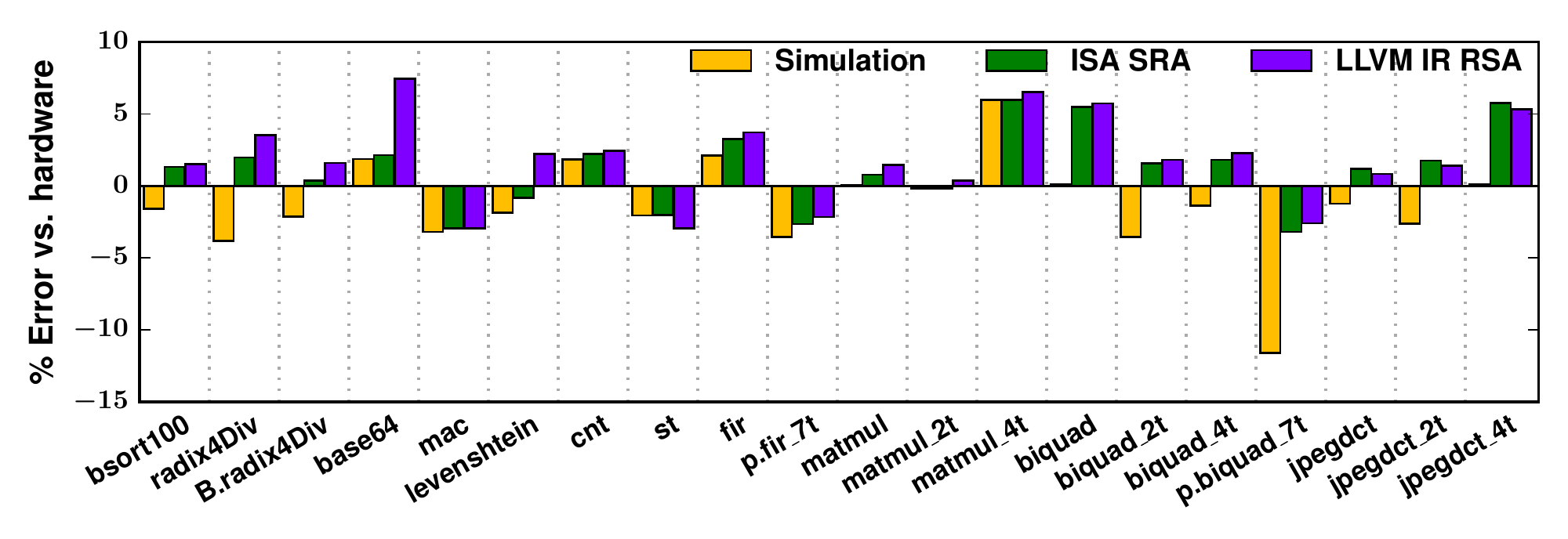}
\caption{SRA and ISS against hardware measurements (worst case program inputs).}
\label{fig:results1}
\end{figure*}

\Cref{fig:results1} presents the error margin of using the same ISA energy model with three energy estimation techniques compared to hardware energy measurements for our benchmarks. \emph{Simulation} produces energy estimations based on instruction traces from ISS, \emph{ISA SRA} uses the model for static analysis at the ISA level and \emph{LLVM IR SRA} uses our mapping technique to apply the model and analysis at LLVM IR level. 

In the case of LLVM IR SRA, the accuracy of the prediction is heavily dependent on the accuracy of the mapping techniques presented in \Cref{sec:mapping}. As shown in \Cref{fig:results1}, for all benchmarks the LLVM IR SRA results are within one percentage point error of ISA SRA results, except for the \texttt{Base64} and \texttt{levenshtein} benchmark with a further 4.3 and 2.0 percentage points error, respectively. In these cases the CFGs of the two levels were significantly different due to new BBs introduced from branches in the ISA level CFG. As discussed in \Cref{subsec:limitations}, substantial differences between the two CFGs can reduce the effectiveness of our current tuning implementation.

Generally, for all results, a proportion of error is present in both forms of static analysis as well as simulation-based energy estimation. The error in the ISS-based estimation is a baseline for the best achievable accuracy in static analysis, as the ISS produces the most accurate execution statistics. For all the benchmarks, the ISA SRA results are over-approximating the trace-based energy estimations. This applies also to the LLVM IR SRA results with exception of the \texttt{st}
benchmark. This over-approximation is a product of the bound analysis used, which is trying to
select the most energy-costly CFG path based on the provided cost model.

The majority of SRA energy estimations are tightly overestimating the actual energy consumption measured on the hardware, up to 5.7\% for ISA SRA and up to 7.4\% for LLVM IR SRA. There are a number of cases for which the retrieved estimations underestimate the actual energy consumption (\texttt{mac}, \texttt{levenshtein}, \texttt{st}, \texttt{p.fir\_7t}, \texttt{p.biquad\_7t}). IPET is intended to provide bounds based on a given cost model. 
In our case it tries to select the worst case execution paths in terms of the energy consumption. 
However, energy consumption is data sensitive, i.e.\ the energy cost of executing an instruction varies, depending on (the circuit switching activity caused by) the operands used. Therefore, the observed underestimation is a consequence of using a data-insensitive energy model and analysis.

The SRA estimations seen in \Cref{fig:results1} represent a loose upper bound on the benchmarks' energy consumption. These bounds cannot be considered safe for use in mission critical applications. However, our experimental results show a low level of SRA underestimation, less than 4\%, and therefore our SRA can still provide valuable guidance to the application programmer, e.g. to compare coding styles or algorithms.

Beyond the use of SRA predictions for bounding the energy consumption, a number of other use cases were investigated and are detailed next.

\subsubsection{SRA alternative use cases}
\label{boundUse}

\begin{figure}[!ht]
    \centering
    \includegraphicsmaybe{width=0.47\textwidth,trim=0.5cm 0.6cm 0.40cm 0.5cm,clip}{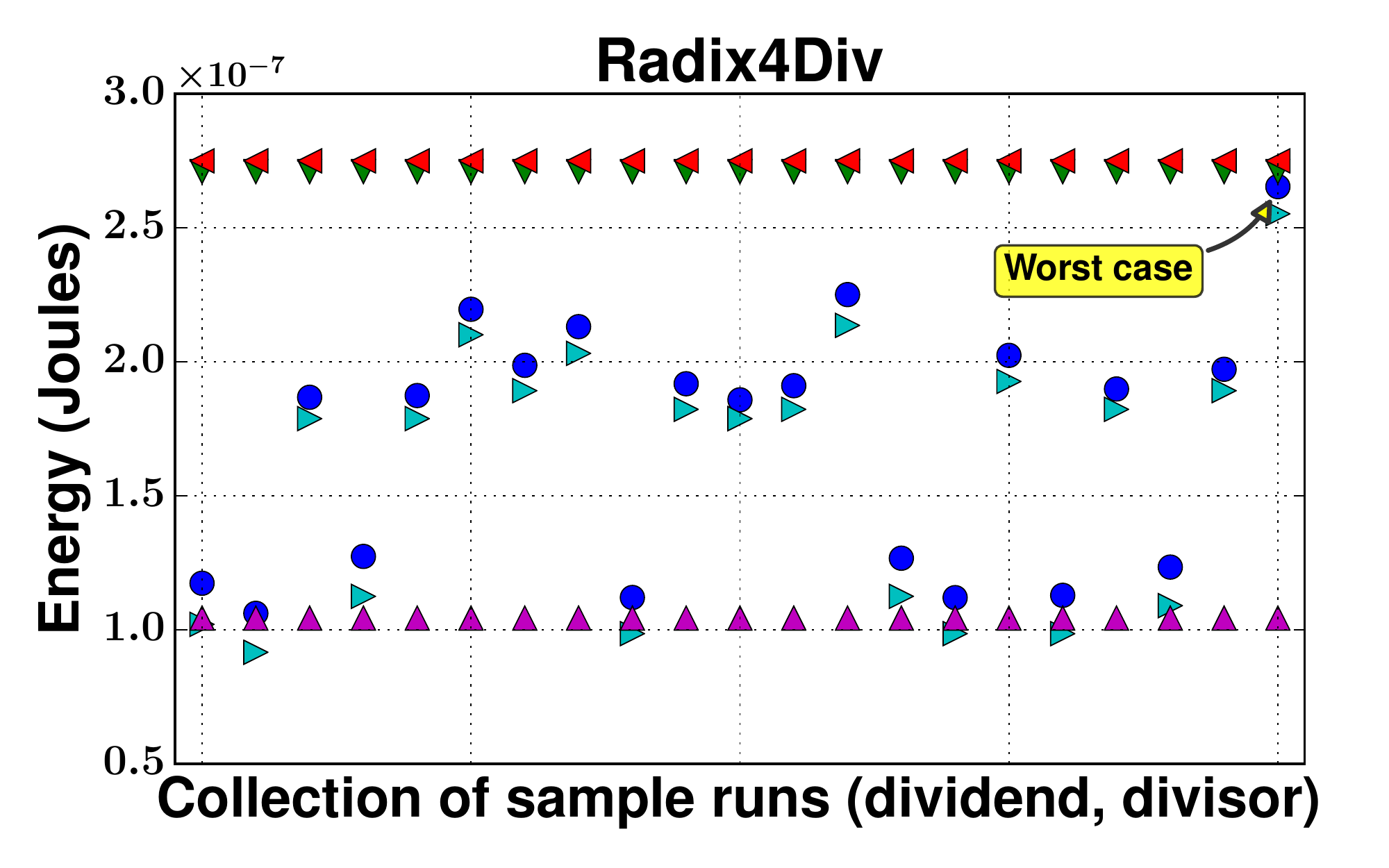}
    \includegraphicsmaybe{width=0.47\textwidth,trim=0.5cm 0.6cm 0.40cm 0.5cm,clip}{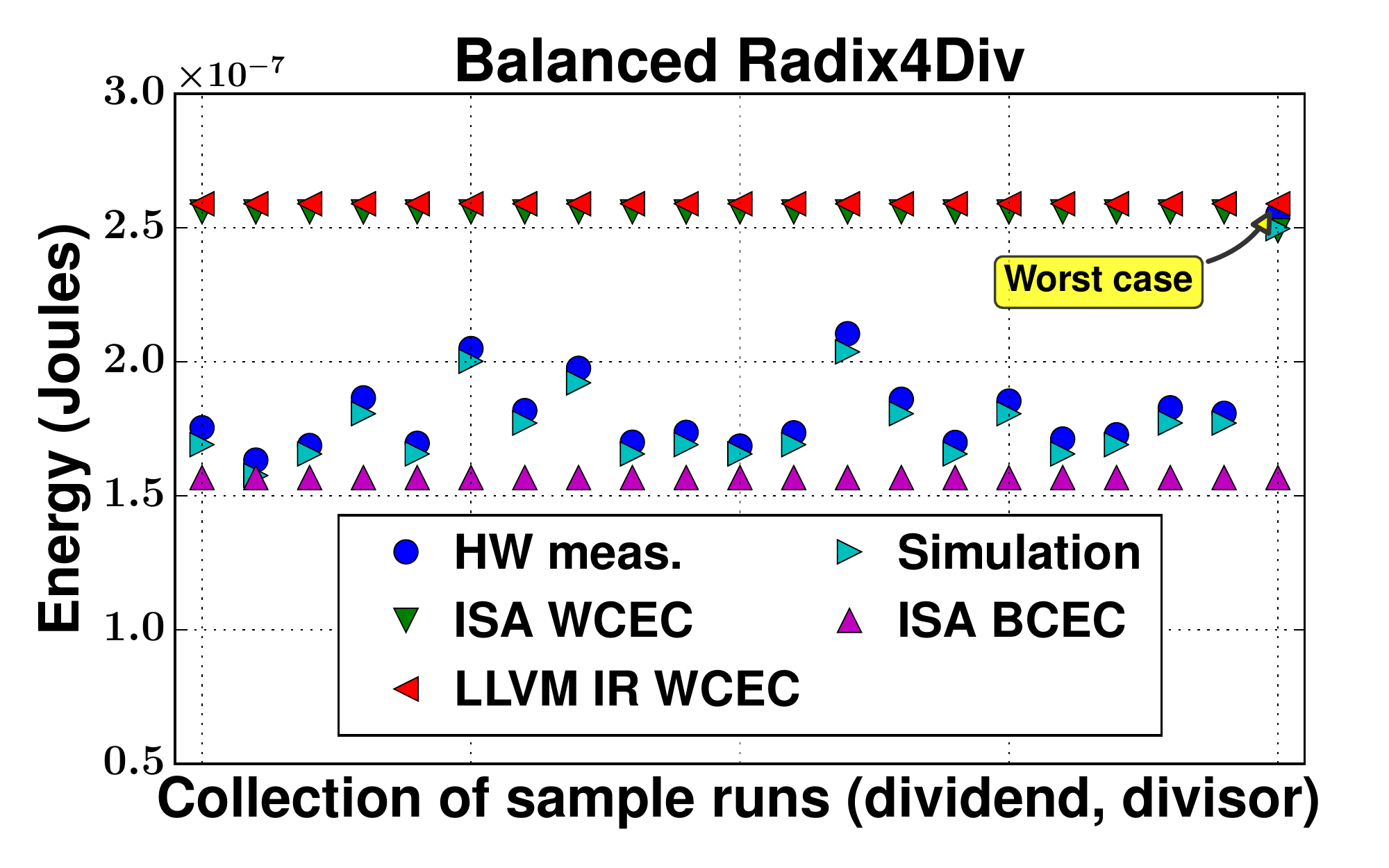}
    \caption{The energy consumption bounds of the optimized and unoptimized version of \texttt{Radix4Div} benchmarks, captured by SRA.}
    \label{fig:results2}
\end{figure}

The modified \texttt{B.Radix4Div} benchmark avoids an early return when the dividend is greater than 255. Omitting this optimization is less efficient, but balances the CFG paths. The effect of this modification can be seen in \Cref{fig:results2}. The ISA level energy consumption lower and upper bounds (the best and worst case retrieved by IPET) are shown. In the optimized version, \texttt{Radix4Div}, the energy consumption across different test cases varies significantly, creating a large range between the upper and lower energy consumption bounds. Conversely, the unoptimized version, \texttt{B.Radix4Div}, shows a lower variation, thus narrowing the margin between the upper and lower bounds, but has a higher average energy consumption.

Knowledge of such energy consumption behavior can be of value for applications like cryptography, where the power profile of systems can be monitored to reveal sensitive information in side channel attacks~\cite{Kocher:1999}. In these situations, SRA can help developers to design code with low energy consumption variation, so that information that could be leaked through power monitoring can be obfuscated.

\subsubsection{Design Space Exploration using SRA}
\label{subsubsect:multi-threaded_usage}

In this section we examine how ISA SRA can be used as an alternative to measurements or simulation for design space exploration of task farms and pipelined programs. We consider applications for which the worst-case execution path is actually the dominant execution path. Having a number of available threads, a number of cores and the ability to apply voltage and frequency scaling, provides a wide range of configuration options in the design phase, with multiple optimization targets. This can include optimizing for quality of service, time and energy, or a combination of all three. For this experiment we used the XMOS XP-SKC-A16~\cite{XMOS_SLICEKIT} development board, which consists of two interconnected Xcores, with 16 threads available in total and supports voltage and frequency scaling.

\begin{figure}[!htb]
\begin{minipage}{\textwidth}
  \begin{minipage}[]{0.3\textwidth}
    \raggedright
    
    \setlength\tabcolsep{1pt}
    \begin{threeparttable}
    \scriptsize
  \begin{tabular}{|c|l|c|c|c|c|}
  \hline
  \rowcolor[HTML]{EFEFEF} 
  \cellcolor[HTML]{EFEFEF} & \multicolumn{1}{c|}{\cellcolor[HTML]{EFEFEF}} & \multicolumn{4}{c|}{\cellcolor[HTML]{EFEFEF}\textbf{Configuration}} \\ \cline{3-6} 
  \rowcolor[HTML]{EFEFEF} 
  \multirow{-2}{*}{\cellcolor[HTML]{EFEFEF}\textbf{Benchmark}} & \multicolumn{1}{c|}{\multirow{-2}{*}{\cellcolor[HTML]{EFEFEF}\textbf{ID}}} & \textbf{C} & \textbf{T} & \textbf{V} & \textbf{F} \\ \hline
   & M1 & 1 & 1 & 1 & 450 \\ \cline{2-6} 
   & M2 & 1 & 2 & 1 & 450 \\ \cline{2-6}
  \multirow{-3}{*}{\begin{tabular}[c]{@{}c@{}}MatMult \\ (4 pairs of\\ 30x30 matrices)\end{tabular}} & M3 & 1 & 4 & 1 & 450 \\ \hline
  & B1 & 1 & 1 & 1 & 450 \\ \cline{2-6} 
  & B2 & 1 & 7 & 0.75 & 150 \\ \cline{2-6} 
  \multirow{-3}{*}{Biquad Filter} & B3 & 2 & 7 & 0.7 & 75 \\ \hline
  \end{tabular}
  \begin{tablenotes}
  \small
  \item \textbf{C:} Number of cores used\\
  \item \textbf{T}: Number of threads used\\
  \item \textbf{V:} Core(s) Voltage in Volts \\
  \item \textbf{F}: Core(s) Frequency in MHz
  \end{tablenotes}
  \end{threeparttable}
\end{minipage}
\hfill
  \begin{minipage}[]{0.54\textwidth}
    \includegraphicsmaybe{width=1\linewidth,trim=0.4cm .5cm 0.6cm 0cm,clip}{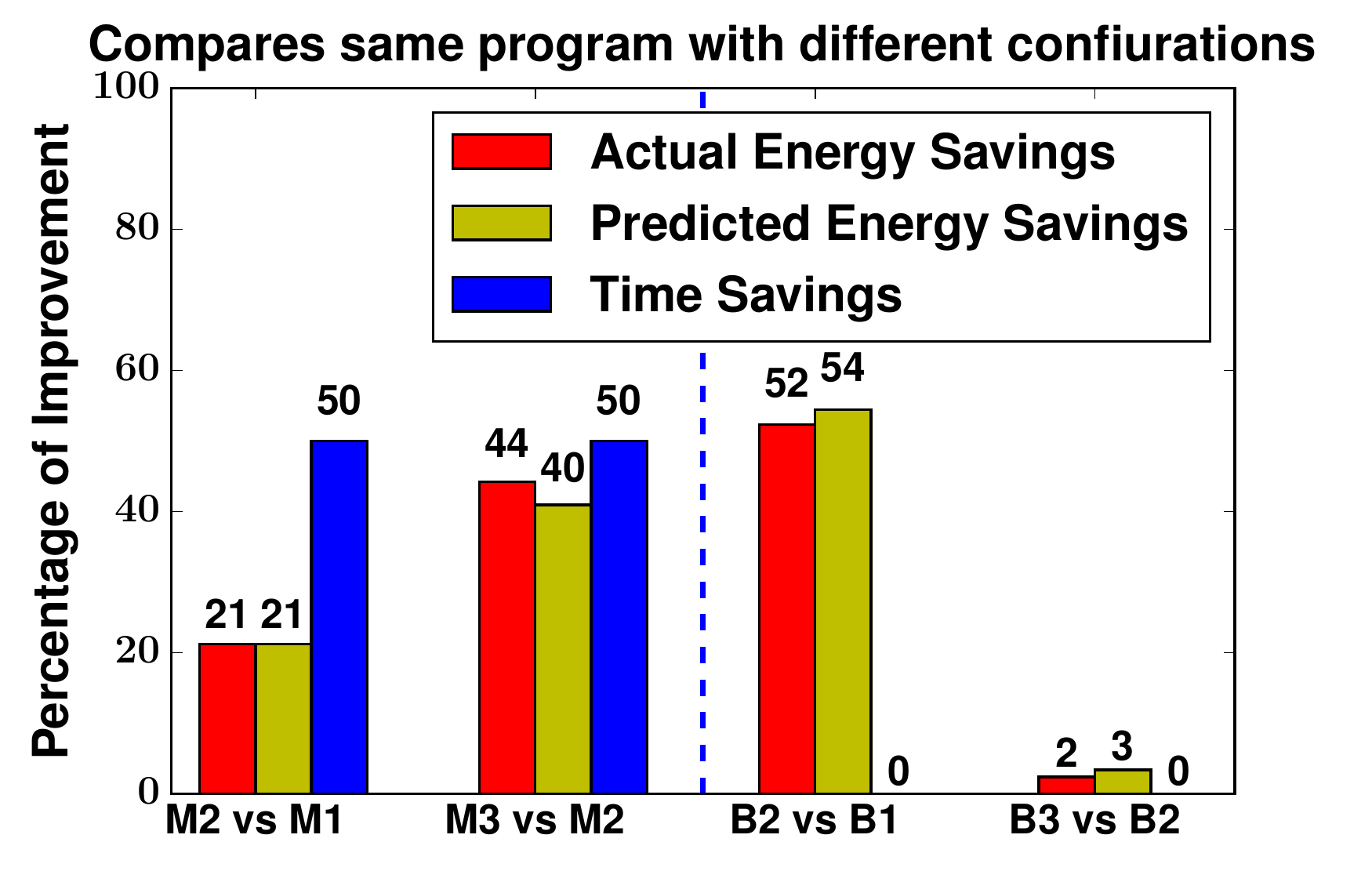}
  \end{minipage}
\caption{Design Space Exploration enabled by SRA.}
\label{fig:DSE}
\end{minipage}
\end{figure}

The left-hand side of \Cref{fig:DSE} shows different configurations for the same program; on the right-hand side these different versions are compared in regards of savings, if any, achieved for time and energy. For energy, both the hardware measurement-based, and the SRA-predicted percentages of savings are shown in order to assess the ability of SRA to support sound energy-aware decisions. Our analysis can infer also the time figures but due to the time-deterministic nature of the architecture the predictions closely match the actual values and therefore we omit them from the graph.

First, SRA was applied to replicated non-communicating threads. The user can make energy-aware decisions on the number of threads to use, with respect to time and energy estimations retrieved by our analysis. As an example, consider four independent matrix multiplications on four pairs of equally sized matrices ($30\times30$). For all three versions we keep the number of cores, the core voltage and frequency constant, but we alter the number of threads and split the computation across them. The single thread version, \texttt{M1}, has an execution time of $4\times$ the time needed to execute one matrix multiplication. However, the two-thread version, \texttt{M2}, halves the execution time and, as indicated by both SRA and actual measurements, decreases the energy by 21\%, compared to \texttt{M1}. The four-thread version, \texttt{M3}, halves the execution time again and, as predicted by SRA, the energy consumption is decreased by 40\% compared to the two-thread version, \texttt{M2}. For the same case, \texttt{M3 vs M2}, the actual energy savings are 44\%. Although there is a different estimation error between different numbers of active threads, the error range of 5\% is small enough to allow comparisons between these different versions, by just using the energy estimations from SRA. The measured and the predicted values are strongly correlated and confirm that using more threads increases the power dissipation, but the reduction in execution time saves energy on the platform under investigation.

Next, SRA was applied to streaming pipelines of communicating threads. There is a choice in how to spread the computation across threads to maximize throughput to either minimize execution time or lower the necessary device operating frequency and voltage while maintaining performance. Our SRA can take advantage of the fact that the energy model used is parametric to voltage and frequency, to statically identify the most energy-efficient configuration of the same program, among a number of options that deliver the same throughput. We demonstrate this on the \texttt{Biquad} benchmark. 

The \texttt{Biquad} benchmark implements an equalizer; it takes a signal and attenuates or amplifies different frequency bands. The \texttt{Biquad} equaliser uses a cascade of biquad filters. Each biquad filter attenuates or amplifies one specific frequency range; the signal is sequentially passed through all filters, eventually attenuating or amplifying all bands. In a standard equalizer setting, a seven-bank filter is used, of which the low four banks are set to amplify, and the top three banks are set to attenuate. These seven banks are implemented in a seven-stage pipeline. 

\Cref{fig:DSE}, on the left-hand side shows the three different versions explored for the
\texttt{Biquad} filter. \texttt{B1} is the sequential version running at 450 MHz and 1V. The
single-core parallel version, \texttt{B2}, uses 7 threads, one for each bank, and runs at 150 MHz and 0.75V. The multi-core parallel version, again using 7 threads, places 4 threads on the first core and 3 on the second core. It therefore allows for halving the core frequency to 75 MHz and a further reduction of the core voltage to 0.7V. All versions, \texttt{B1}, \texttt{B2} and \texttt{B3} deliver the same filter performance, therefore the percentage of time savings on the right hand side graph of \Cref{fig:DSE} is zero. In contrast, we can see a predicted energy saving on \texttt{B2 vs B1} of 54\% and on \texttt{B3 vs B2} of 3\%. The actual energy savings are 52\% and 2\%, respectively. The small error of our energy consumption estimations enables energy-aware decisions just by the use of SRA. In this case, version \texttt{B2} is the best solution since, while maintaining the same performance, it halves the energy consumption. In comparison, the small additional energy savings realized by \texttt{B3} may not warrant the use of a second core. These energy-aware decisions can not be achieved by only examining the execution time of each one of the three configurations. Similar results were achieved for the \texttt{Fir} benchmark.

In both of the above examples, we demonstrated that our SRA is sufficient to provide design space exploration guidance considering both time and energy, without the need of any simulation or hardware measurements. This has significant benefits since SRA is faster than simulation and less costly to deploy than hardware measurements.

\subsection{Profiling-Based Analysis Results}

\begin{figure*}[ht]
\centering
\includegraphics[width=1\textwidth,trim=0.3cm 0.45cm 0cm 0.40cm,clip]{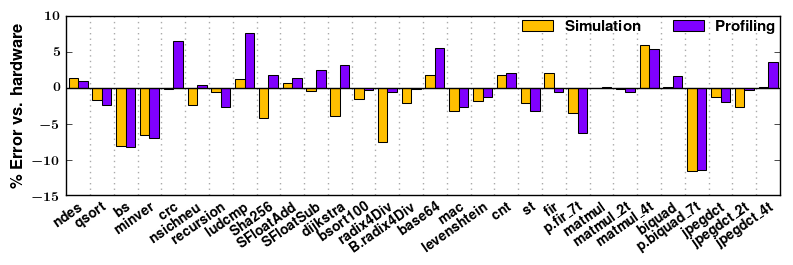}
\caption{Profiling- and ISS-based energy estimations against hardware measurements.}
\label{fig:ProfResults}
\end{figure*}

To evaluate the profiling-based energy estimation technique together with the mapping technique, 30 benchmarks were used. For these benchmarks, \Cref{fig:ProfResults} presents the error margin of using the same ISA energy model with the ISS-based estimations and the profiling-based estimations, compared to hardware energy measurements, respectively. The average absolute error obtained for the profiling-based estimations is 3.1\%, and 2.7\% for the ISS-based estimations. The ISS-based estimation is more reliable than profiling. This is because cycle-accurate ISS allows for a very precise energy consumption estimation for each particular program execution as the program is executed in simulation with a specific set of input parameters. The exact sequence of instructions can be recorded during simulation and then used to estimate energy consumption. In contrast, profiling is less precise as it operates further away from the hardware, at the LLVM IR level. Profiling estimation precision heavily depends on the quality of the mapping techniques introduced in~\Cref{subsec:mapXcore}, and on the quality of the profiling techniques used to collect execution counts of LLVM IR BBs, as described in \Cref{subsec:Profiling}. However, our profiling results demonstrate a high accuracy with an average error deviation of 1.8\% from the ISS.

\subsubsection{Profiling-Based Estimation Performance}

ISS execution is often several orders of magnitude slower than hardware. The performance of an ISS is governed by the complexity of the program's underlying algorithms. In contrast, the performance of our profiling is mainly governed by the program size, since retrieving the BB execution counts incurs a negligible execution time overhead when running on the actual platform, as discussed in \Cref{subsec:Profiling}. The bigger the program the more time will be needed for Stage 1, \texttt{LLVM Instrumentation Pass} in the \texttt{Profiling} phase and Stage 1, \texttt{BB Energy Characterization}, in the \texttt{Energy Consumption Estimation} phase, shown in \Cref{fig:ProfilerOverview}. 

In the case of benchmarks with low algorithmic complexity and function inputs that will trigger a
very short simulation time, no significant performance gains will be observed from using profiling over ISS estimation. For example, for our \texttt{SFloatAdd} benchmark, with a $O(1)$ complexity, the average profiling estimation performance observed has a negligible gain over the ISS estimation. On the other hand, increasing the complexity of the benchmarks' underlying algorithms and using parameters which trigger longer simulation time, results in the profiler significantly outperforming the simulation. For example, when using $30\times30$ size matrices for our matrix multiplication benchmark, the profiling estimation achieves a 381 times speedup over the ISS estimation. The benchmark's small size allowed for a fast profiling estimation, but the ISS estimation performance follows the $O(n^3)$ algorithmic complexity of the benchmark.

A further speedup can be achieved for the profiling-based method, when testing the same program with different inputs. If the optimized LLVM IR code obtained is the same across the different inputs, then there is no need to repeat the LLVM IR energy characterization stage. In such cases using the profiling energy estimation has clear performance advantages compared to the ISS-based energy estimation.

\subsection{Discussion}

Our SRA-based estimation at the LLVM IR level has a deviation in the range of 1\% from the ISA SRA, and our profiling-based estimation has an average error deviation of 1.8\% from the ISS. This shows that the tuning phase introduced in \Cref{subsub:MapTuning}, mitigates the impact of the mapping limitations described in \Cref{subsec:limitations}, and yields sufficiently accurate results for the architecture and compiler under consideration. Depending on the architecture and the compiler implementation, the tuning phase heuristics can be further improved to achieve the required level of accuracy.

\section{Related Work}
\label{sec:background}

SRA is a methodology to determine usage bounds of a resource (usually time or energy or both) for a specific task when executed on a piece of hardware, without actually executing the task. This requires accurate modeling of the hardware in order to capture the dynamic
functional and non-functional properties of task execution. Determining these properties accurately is known to be undecidable in general. Therefore, to extract safe values for the resource usage of a task, a sound approximation is needed~\cite{Wilhelm:2008,brat2014ikos}.

SRA has been mainly driven by the timing analysis community. Static cost analysis techniques based on setting up and solving recurrence equations date back to Wegbreit's~\cite{Wegbreit:1975} seminal paper, and have been developed significantly in subsequent work~\cite{Rosendahl:1989,Debray:1997,Vasconcelos:2003,Navas:2007,Albert2011}. Other classes of approaches to cost analysis use dependent types~\cite{DBLP:journals/toplas/0002AH12}, SMT solvers~\cite{Alonso2012}, or size change abstraction~\cite{DBLP:journals/corr/abs-1203-5303}. Generally, for performing an accurate WCET static analysis, there are four essential components~\cite{Wilhelm:2008}:

\begin{enumerate}
 \item Value analysis: mainly used to analyze the behavior of the data cache.
\item Control flow analysis: used to identify the dynamic behavior of a program.
\item Low-level analysis: attempts to retrieve timing costs for each atomic unit on a given hardware platform, such as an instruction or a BB in a CFG for a processor.
\item Calculation: uses the results from the two previous components to estimate the WCET. The most common techniques used for calculation of the WCET are the IPET, the path-based techniques and the tree-based methods~\cite{Engblom:2000a}.
\end{enumerate}

Three of the above components, namely the control flow analysis, low-level analysis and calculation, are adopted in our work, as discussed in \Cref{sec:SAnalysis}.

IPET is one of the most popular methods used for WCET analysis~\cite{Li:1995b,Theiling:1998,Ottosson:1997,Engblom:2000a,PotopButucaru:2013}. In this approach, the CFG of a program is expressed as an ILP system, where the objective function represents the program's execution time. The problem then becomes a search for the WCET by maximizing the retrieved objective function under some constraints on the execution counts of the CFG's BBs.

Although significant research has been conducted in static analysis for the execution time estimation of a program, there is little on energy consumption. One of the few approaches~\cite{Navas:2008} seeks to statically infer the energy consumption of Java programs as functions of input data sizes, by specializing a generic resource analyzer~\cite{Navas:2007,Hermenegildo:2005} to Java bytecode analysis~\cite{Navas:2009}. However, a comparison of the results to actual
measurements was not performed.
Later, in \cite{Liqat:2014}, the same generic resource analyzer was instantiated to perform energy analysis of XC~programs~\cite{Watt:2009} at the ISA level based on ISA-level energy models and including a comparison to actual hardware measurements. However, the scope of this particular analysis approach was limited to a small set of simple benchmarks because information required for the analysis of more complex programs, such as program structure and types, is not available at the ISA level. The SRA presented in this paper does not rely on such information. 
A similar approach, using cost functions, was used in \cite{grech15}. The analysis was performed at the LLVM IR level, using an early version of the mapping technique that we formalize and describe in full detail for the first time in this paper. Although the range of programs that could be analyzed was improved, compared to \cite{Liqat:2014}, the complexity of solving recurrence equations for analyzing larger programs proved a limiting factor.

In \cite{Jayaseelan2006} the WCEC for a program was inferred by using the IPET first introduced in \cite{Li:1995b}. Although the authors manage to bound the WCEC on a simulated processor by maximizing the switching activity factor for each simulated component, they acknowledge the need to validate their estimation results against commercial embedded processors. Similarly, in \cite{wagemannworst} the authors attempt to perform static WCEC analysis for a simple embedded processor, the \texttt{ARM Cortex M0+}. This analysis is also based on IPET combined with a so-called absolute energy model, an energy model that is said to provide the ``maximum energy consumption of each instruction''. The authors argue that they can retrieve a safe bound. However, this is demonstrated on a single benchmark, {\tt bubblesort}, only. They also acknowledge that using an absolute energy model can lead to significant overestimation, making the bounds less useful. We choose to use a pseudo-random data characterized energy model, as empirical evidence shows that such models tend to be close to the actual worst case~\cite{Pallister:15}.

All of the reviewed previous works combined SRA techniques together with energy models to capture the WCEC path. Currently, there is no practical method to perform average case static analysis~\cite{townley2013practical}. One of the most recent works towards average case SRA~\cite{Schellekens201061} demonstrates that compositionality combined with the capacity for tracking data distributions unlocks the average case analysis, but novel language features and hardware designs are required to support these properties. This situation motivated us to develop a dynamic profiling technique that captures the actual energy consumption of a program based on its input parameters.

There are two main requirements for CPU energy profiling. First, a technique is needed to instrument a program's execution and collect data that represent the program's dynamic behavior. 
Code instrumentation, simulation and hardware performance counters are some of the most popular techniques used to collect such data. The second requirement is a method of associating energy information with the various entities, events in the set of the collected instrumentation data. This is typically done either through an energy model, or by power-monitoring the program's execution.

Energy modeling and profiling can be performed at various levels of abstraction. In \cite{Bogliolo1997}, a gate-level power consumption model was created and combined with event-driven logic simulation to estimate the power consumption of programs. Modeling and profiling at such low design levels is not practical for most commercial embedded processors since their lower level circuit information is not available. Moreover, this estimation is quite slow and not practical for fine-grained energy characterization of software.

In \cite{Tiwari1996} an ISA-level energy model was proposed which treated the hardware as a black-box and obtained energy consumption data through hardware measurements of large loops of individual instructions. Modeling at the ISA allowed for attributing energy costs to low-level software components such as ISA BBs. This led to further research that combined ISA energy models with instruction set simulators and profilers to extract energy estimations~\cite{Sarta1999,Brooks2000,Steinke2001,Sami2002,Ibrahim2008}. Our energy model is also based on \cite{Tiwari1996}, but significantly extended to a multi-threaded version for the Xcore, which takes into account both inter-instruction and inter-thread effects.

For more complex processors or for system-level energy consumption estimation, energy modeling and profiling at the ISA level is impractical. In such cases, performance counter-based statistical energy modeling and estimation is preferable. In several works~\cite{xscaleunitevents,phimodel,Schubert:2012}, the authors used performance counters to characterize the processor energy consumption based on the conditions affecting these counters, such as cache misses and prefetches. Then, by combining this energy characterization with performance counters execution statistics, they predict the energy consumption of an application. An alternative to performance counters is the use of an external multi-meter to directly measure a system's energy profile~\cite{Weaver:2012}. 

In \cite{Brandolese2011}, energy characterization of LLVM IR code is performed by linear regression analysis. This is combined with instrumentation and execution on a target host machine to estimate the performance and energy requirements in embedded software. Transferring the LLVM IR energy model to a new platform requires performing the regression analysis again. The mapping technique we present here does not involve any regression analysis and it is fully portable. It requires only the adjustment of the LLVM mapping pass to the new architecture. Furthermore, our LLVM IR mapping technique provides an on-the-fly energy characterization that takes into consideration the compiler behavior including optimization and transformation passes and other architecture-specific aspects. The instrumentation technique of \cite{Brandolese2011} is also based on collecting execution BB counts, but this is done by executing programs on a host machine rather than the actual platform. However, compiling and executing on a higher-end PC, rather than the target deeply embedded device, could yield an execution profile significantly different from the one that would be obtained on the target machine. Our profiling technique collects execution counts on the target machine, and maps them back to their corresponding LLVM IR BB.

In \cite{Ozturk:2013}, a scheme is proposed that enables the compiler to exploit both task and data parallelism and automatically map an application to an embedded multi-processor containing voltage islands. The scheme is based on the workload of each processor, using both hardware parallelism and voltage scaling to reduce energy consumption without increasing the overall execution time. Per-core frequency and voltage configurations can be provided to appropriately design Xcore-based systems. Using our profiling technique and our LLVM IR energy characterization, accurate workload estimations can be achieved at the LLVM IR level. These could be utilized by the optimization scheme introduced in \cite{Ozturk:2013} to enable the LLVM compiler to achieve similar energy consumption savings.

Profiling has previously been used successfully to enable energy-aware compilation. A combination of code analysis and profiling techniques can enable the compiler to build power-aware flow graphs~\cite{Rele2002}. Such a graph has its basic blocks annotated with the hardware resource requirements and the execution counts. The power-aware flow graphs can then be used to identify code regions where several of the processor's functional units can be turned off during execution to reduce the static power dissipation.~\cite{Lin:2015} demonstrate a framework that enables the compiler to support a number of energy-related optimizations for parallel design patterns. A series of pragmas were introduced to identify specific parallel design patterns and guide the compiler to apply power optimizations. These optimizations were enabled by utilizing power profiling and code instrumentation feedback provided by a simulator.

Our profiler provides energy consumption information directly into the compiler's optimizer. Therefore it can enable feedback-directed compilation that targets energy-specific optimizations and design space exploration for deeply embedded systems. 

\section{Conclusion and Future Work}
\label{sec:conc_future}

This work focuses on providing energy consumption estimation techniques that can enable energy transparency in the software stack of deeply embedded processors, typically used in IoT applications. A novel target-agnostic mapping technique is introduced to allow energy characterization of the LLVM tool chain intermediate representation, namely the LLVM IR. This is a significant step beyond existing ISA-level energy estimation techniques. Our mapping technique 
can give powerful insights to the LLVM IR optimizer regarding execution time and the energy consumption of a program. This enables programmers to investigate how optimizations can affect their program's energy consumption, or even help introduce new energy-specific optimizations in future.

For energy consumption bounds an IPET-based SRA was developed. The analysis was introduced at both ISA and LLVM IR levels. The results demonstrated that the mapping technique enables SRA at the LLVM IR level with a small accuracy loss in the range of only 1\%, compared to SRA at ISA level. SRA-based energy estimation was also applied to a set of multi-threaded programs for the first time to our knowledge. This is a significant step beyond existing work that examines single-thread programs. As shown in \Cref{subsubsect:multi-threaded_usage}, such an analysis can provide significant guidance for time-energy design space exploration between different numbers of threads and cores.

To estimate the actual energy consumption of a program under specific input parameters, a target-agnostic profiling technique was developed. The technique is enabled by the LLVM IR energy characterization utilizing our mapping. It is designed to ensure that the instrumentation code required for profiling does not lead to energy overheads. Experimental evaluation shows an average absolute error of 3.2\% compared to hardware measurements. Our profiling technique is more flexible and significantly more efficient than ISS-based estimation. It can attribute energy consumption to software components, such as BBs and functions, in contrast to hardware measurements. The high accuracy and performance of our profiler can enable feedback-directed optimization for energy consumption. LLVM IR energy consumption estimations of each compilation can be taken into consideration iteratively in subsequent compilations. Each new compilation will be able to make more energy-aware decisions.

The techniques introduced in this paper are focused on deeply embedded architectures that favor predictability over performance. Such architectures are the backbone of IoT applications. Enabling energy-aware software development for such architectures will help to tackle the energy challenge that IoT faces. 
We acknowledge that some of our techniques, mainly SRA-based estimation, work best with predictable architectures. This is also true for WCET analysis. In fact, static analysis is inherently limited in that sense. Thus, in future work we intend to examine the portability of our profiling-based techniques to more complex processors.

Future work aims to analyze more complex concurrent programs, such as distinct non-communicating threads, or pipelines of threads with unbalanced workloads. We anticipate that our LLVM IR profiling technique will scale better to these cases than SRA. As the core-count of the analyzed system grows, more in-depth exploration of energy saving techniques such as per-core DVFS could be
considered both at the model level and SRA or profiling levels.

\section*{Acknowledgments}\label{sec:Acknowledgments}
  The research leading to these results has received funding from the European
  Union Seventh Framework Programme (FP7/2007-2013) under grant agreement no.\
  318337, ENTRA: Whole-Systems Energy Transparency, and from the ARTEMIS Joint
  Undertaking under grant agreement no.\ 621429 (project EMC2).

\bibliography{typeinst}

\end{document}